\documentclass[a4paper,11pt]{article}
\usepackage{pos}
\usepackage{amsmath}
\usepackage{graphicx}
\usepackage{booktabs}
\usepackage{bm}

\title{Hyperon--nucleon interactions at BESIII}

\author*[a]{Hong-Fei Shen}
\onbehalf{on behalf of the BESIII Collaboration}

\affiliation[a]{Department of Physics, Indiana University Bloomington,\\
Bloomington, Indiana 47405, USA}

\emailAdd{shenhongfei@ihep.ac.cn}
\emailAdd{shenho@iu.edu}

\abstract{
    Hyperon--nucleon interactions provide essential information on the
strange sector of baryon--baryon forces, with implications for
hypernuclei, dense baryonic matter, and non-perturbative QCD.
Direct experimental information, however, remains scarce because
hyperons are short lived and intense conventional hyperon beams are
difficult to produce.
At BESIII, hyperons and antihyperons produced in charmonium decays
can instead be used as incident particles, while the detector material
serves as an effective fixed target.
This approach has enabled a new program of hyperon--nucleon studies
at an $e^+e^-$ collider.
Representative results include the first study of
$\Xi^{0}n\to\Xi^-p$ using $\Xi^0$--nucleus scattering,
measurements of elastic $\Lambda p\to\Lambda p$ and
$\bar{\Lambda}p\to\bar{\Lambda}p$ scattering, and recent observations
of $\bar{\Lambda}p$ annihilation into multi-meson final states.
These measurements demonstrate that BESIII can operate as an
effective hyperon--nucleon scattering facility and provide new
experimental constraints on the poorly explored strange sector of
baryon--baryon interactions.
}

\FullConference{43rd International Conference on High Energy Physics (ICHEP2026)\\
 July 30, 2026 to August 5, 2026\\
Natal, Brazil\\}

\begin{document}
\maketitle

\section{Introduction}

The interaction between baryons is one of the central ingredients in
our understanding of strongly interacting matter.
While the nucleon--nucleon interaction has been extensively studied,
introducing strangeness extends the problem to the
hyperon--nucleon ($YN$) and hyperon--hyperon sectors.
Experimental information on these interactions is important in several
contexts.

In hypernuclei, $YN$ interactions determine binding energies and level
structures and therefore provide direct information on the nuclear
force in the strange-quark sector.
At high baryon densities, hyperons may also appear in the interior of
neutron stars, where hyperon--nucleon and hyperon--hyperon interactions
can affect the equation of state.
Finally, $YN$ scattering data provide valuable constraints on
baryon--baryon potentials, chiral effective-field-theory approaches,
and lattice-QCD calculations.

In contrast to the abundant information available for nucleon--nucleon
scattering, direct $YN$ scattering measurements remain limited.
The main experimental difficulty is the short lifetime of hyperons.
Conventional fixed-target scattering experiments require sufficiently
intense incident particle beams, whereas hyperon beams are difficult
to produce and transport.
Measurements involving antihyperons are even more challenging.

\begin{figure}[t]
  \centering
  \includegraphics[width=8cm]{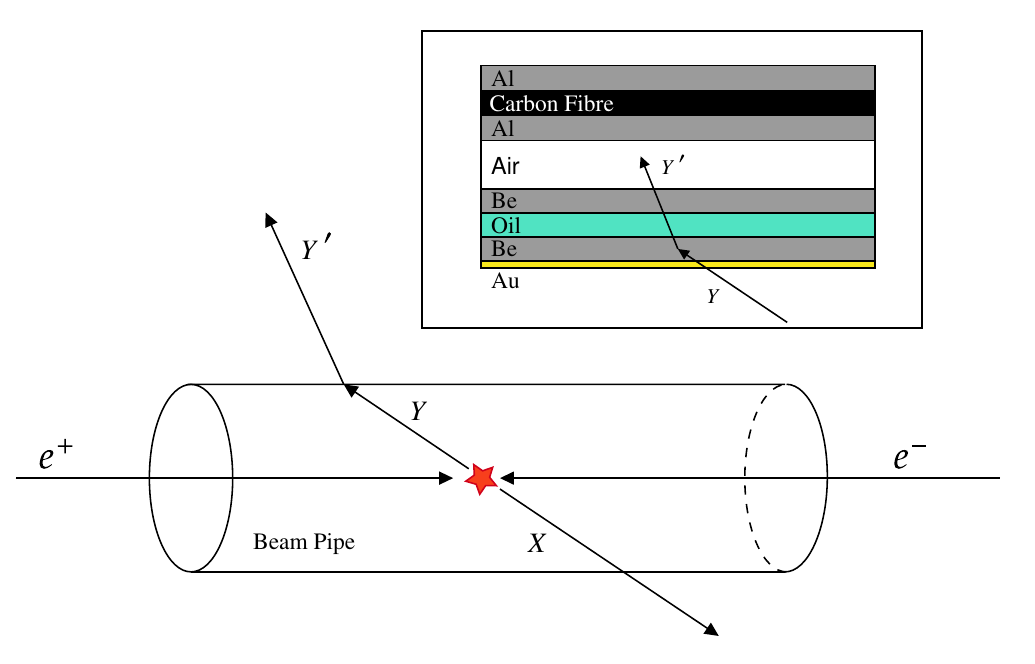}
  \caption{%
  Schematic illustration of the hyperon--nucleon interaction method
  at BESIII. Hyperons produced in charmonium decays serve as incident
  particles, while nuclei in the detector material provide an
  effective fixed target.
  }
  \label{fig:method}
\end{figure}

BESIII provides a complementary approach to this problem.
Large charmonium data samples contain abundant two-body decays to
hyperon--antihyperon pairs.
The resulting hyperons are nearly monoenergetic and may interact with
the material surrounding the $e^+e^-$ interaction point before
decaying.
Consequently, the detector material itself can be exploited as an
effective fixed target~\cite{Dai2024}.
The basic concept of this approach is illustrated in Fig.~\ref{fig:method}.

\section{BESIII as a hyperon--nucleon scattering facility}

BESIII has collected approximately
$10\times10^{9}$ $J/\psi$ events and
$2.7\times10^{9}$ $\psi(3686)$ events~\cite{BESIII:2021cxx, BESIII:2024lks}.
Charmonium decays provide clean sources of
$\Lambda$, $\Sigma$, and $\Xi$ hyperons and their antiparticles~\cite{BESIII:2022qax, BESIII:2025jxt, BESIII:2024nif, BESIII:2023drj, BESIII:2026hgj, Li:2016tlt}.
For two-body decays, the incident hyperon momentum is narrowly
distributed and is determined by the decay kinematics.

The material close to the interaction point provides several possible
nuclear targets.
In particular, the beam pipe contains layers of gold, beryllium, and
cooling oil, while additional material is present in the inner wall of
the main drift chamber.
Hydrogen in the beam-pipe cooling oil is especially useful because it
provides nearly free protons at rest.
This makes possible direct studies of hyperon--proton reactions with
well-defined initial-state kinematics.

For a detector layer with density $\rho_i$ and molar mass $M_i$, the
number density of target atoms is
\begin{equation}
  n_i = \frac{N_{\rm A}\rho_i}{M_i},
  \label{eq:numberdensity}
\end{equation}
where $N_{\rm A}$ is Avogadro's constant.
Combining the target density with the incident-hyperon flux and its
path length through each material layer defines an effective target
luminosity.
Schematically, the measured cross section can be expressed as
\begin{equation}
  \sigma =
  \frac{N_{\rm sig}}
       {\epsilon\,{\cal B}\,{\cal L}_{\rm eff}},
  \label{eq:crosssection}
\end{equation}
where $N_{\rm sig}$ is the signal yield, $\epsilon$ is the detection
efficiency, ${\cal B}$ denotes the relevant decay branching fractions,
and ${\cal L}_{\rm eff}$ is the effective luminosity.
The detector material is therefore treated not simply as passive
material or a source of background, but as a target with known
geometry and composition.

\section{Hyperon--nucleon scattering measurements}

\subsection{$\Xi^0 n\to\Xi^-p$}

The first application of this method at BESIII was the study of
\begin{equation}
  \Xi^0 n \to \Xi^- p,
\end{equation}
using $\Xi^0$ hyperons produced in
$J/\psi\to\Xi^0\bar{\Xi}^0$ decays~\cite{BESIII2023Xi}.
The $\Xi^0$ momentum is approximately
$0.818~{\rm GeV}/c$, and the target neutron is contained in the
$^{9}{\rm Be}$, $^{12}{\rm C}$, or $^{197}{\rm Au}$ nuclei of the
beam-pipe material.

\begin{figure}[t]
  \centering
  \includegraphics[width=6cm]{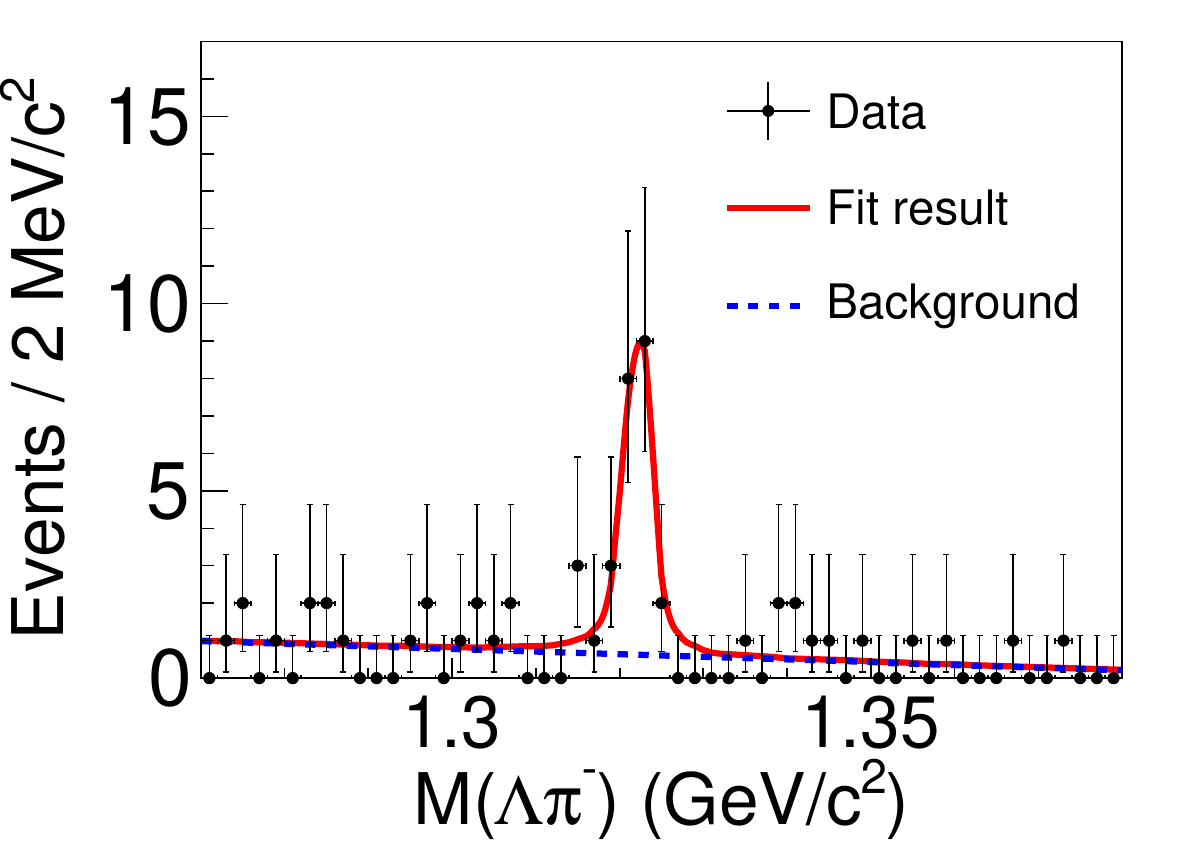}
  \caption{%
      Distribution of $M(\Lambda\pi^-)$ in data (dots with error bars).
  }
  \label{fig:Xiscattering}
\end{figure}

A clear signal is observed with a statistical significance of
$7.1\sigma$, as shown in Fig.~\ref{fig:Xiscattering}.
For the $^{9}{\rm Be}$ target, the measured cross section is
\begin{equation}
\begin{split}
 \sigma\bigl(
 \Xi^0+{}^{9}{\rm Be}
 \to\Xi^-+p+{}^{8}{\rm Be}
 \bigr)
 =
 (22.1\pm5.3_{\rm stat}\pm4.5_{\rm syst})~{\rm mb}.
\end{split}
\end{equation}
No significant $H$-dibaryon signal is found in the $\Xi^-p$ final
state.
This measurement represents the first study of a hyperon--nucleon
interaction using hyperons produced at an $e^+e^-$ collider.

\subsection{Elastic $\Lambda p$ and $\bar{\Lambda}p$ scattering}

An especially clean environment is obtained from
\begin{equation}
  J/\psi\to\Lambda\bar{\Lambda},
\end{equation}
which produces $\Lambda$ and $\bar{\Lambda}$ baryons with a momentum
of approximately $1.074~{\rm GeV}/c$.
When they interact with nearly static hydrogen nuclei in the
beam-pipe cooling oil, the corresponding center-of-mass energy is
approximately
\begin{equation}
  \sqrt{s}\simeq2.243~{\rm GeV}.
\end{equation}

\begin{figure}[t]
  \centering
  \includegraphics[width=6cm]{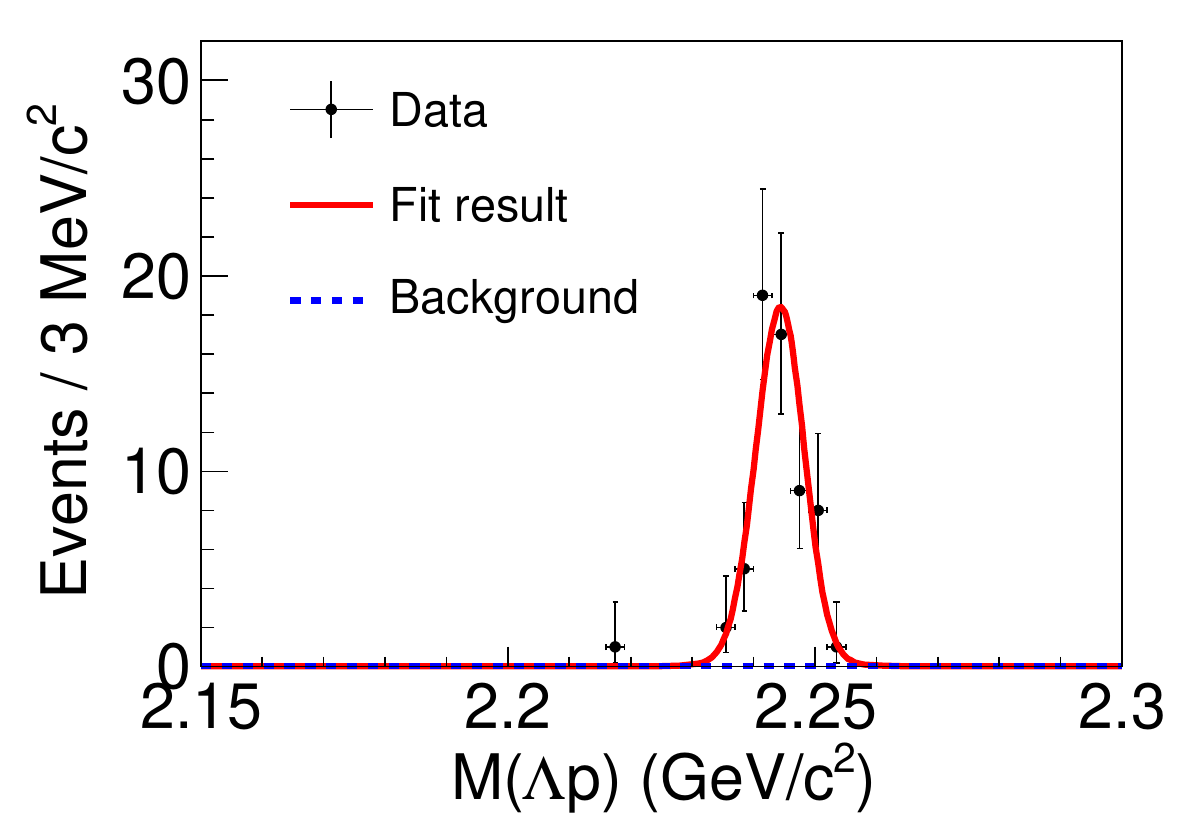}
  \includegraphics[width=6cm]{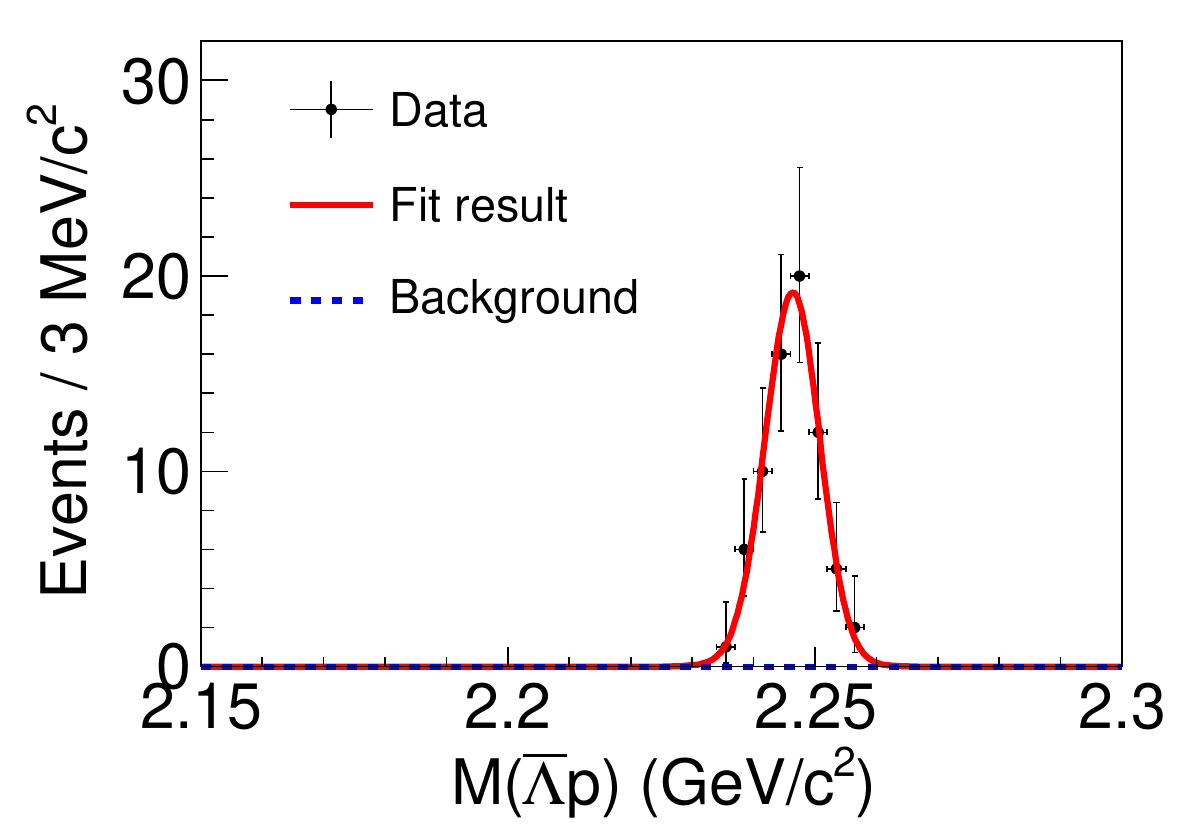}
  \caption{Distributions of $M(\Lambda p)$ (left) and $M(\bar{\Lambda}p)$ (right) in data (black dots with error bars) for the reactions $\Lambda p\to\Lambda p$ and $\bar{\Lambda}p\to\bar{\Lambda}p$, respectively.}
  \label{fig:Lamscattering}
\end{figure}

Clear signals of
$\Lambda p\to\Lambda p$ and
$\bar{\Lambda}p\to\bar{\Lambda}p$
are observed~\cite{BESIII2024Elastic}, as shown in Fig.~\ref{fig:Lamscattering}.
Within the fiducial region
$-0.9\leq\cos\theta\leq0.9$, the cross sections are measured to be
\begin{align}
 \sigma(\Lambda p\to\Lambda p)
 &= (12.2\pm1.6_{\rm stat}\pm1.1_{\rm syst})~{\rm mb},\\
 \sigma(\bar{\Lambda}p\to\bar{\Lambda}p)
 &= (17.5\pm2.1_{\rm stat}\pm1.6_{\rm syst})~{\rm mb}.
\end{align}
The $\bar{\Lambda}p$ measurement constitutes the first study of
antihyperon--nucleon elastic scattering.

The differential cross sections display markedly different angular
behavior.
The $\Lambda p$ reaction shows only a mild tendency toward forward
scattering, whereas the $\bar{\Lambda}p$ distribution exhibits a
pronounced forward enhancement.
An effective-Lagrangian calculation based on one-boson exchange
provides a possible interpretation~\cite{Wang2025}.
In this description, both $t$- and $u$-channel contributions are
allowed for $\Lambda p$ elastic scattering, whereas the
$\bar{\Lambda}p$ process receives only the $t$-channel contribution,
naturally enhancing the forward region.

\section{Antihyperon--nucleon annihilation}

The same nearly monoenergetic $\bar{\Lambda}$ source and hydrogen
target can also be used to study annihilation into mesons.
Using the full $J/\psi$ sample, BESIII studied the reactions
\begin{equation}
 \bar{\Lambda}p
 \to K^+\pi^+\pi^- + k\pi^0,
 \qquad k=1,2,3,
\end{equation}
at an incident $\bar{\Lambda}$ momentum of approximately
$1.074~{\rm GeV}/c$~\cite{BESIII2026Annihilation}.

The channels
\begin{align}
 \bar{\Lambda}p
 &\to K^+\pi^+\pi^-\pi^0,\\
 \bar{\Lambda}p
 &\to K^+\pi^+\pi^-2\pi^0
\end{align}
are observed for the first time.
Their cross sections are
\begin{align}
 \sigma_{\bar{\Lambda}p\to K^+\pi^+\pi^-\pi^0}
 &=
 8.5^{+1.2}_{-1.1}\,(\mathrm{stat})
 \pm0.4\,(\mathrm{syst})~{\rm mb},\\
 \sigma_{\bar{\Lambda}p\to K^+\pi^+\pi^-2\pi^0}
 &=
 7.9^{+1.9}_{-1.7}\,(\mathrm{stat})
 \pm0.4\,(\mathrm{syst})~{\rm mb}.
\end{align}
No significant signal is found for the $k=3$ final state, for which
an upper limit of $7.2~{\rm mb}$ is obtained at 90\% confidence level.

Evidence for a $K^{*}(892)^+$ contribution is also found in the
$K^+\pi^0$ invariant-mass distribution of the $k=1$ channel.
The measured cross section for
$\bar{\Lambda}p\to K^{*}(892)^+\pi^+\pi^-$ is
\begin{equation}
 \sigma =
 12.5^{+3.8}_{-3.4}\,(\mathrm{stat})
 \pm1.2\,(\mathrm{syst})~{\rm mb},
\end{equation}
where possible interference effects are not included because of the
limited statistics.
These results provide direct experimental information on
antihyperon--nucleon annihilation dynamics in a region where
experimental constraints have been very limited.

\begin{figure}[t]
  \centering
  \includegraphics[width=10cm]{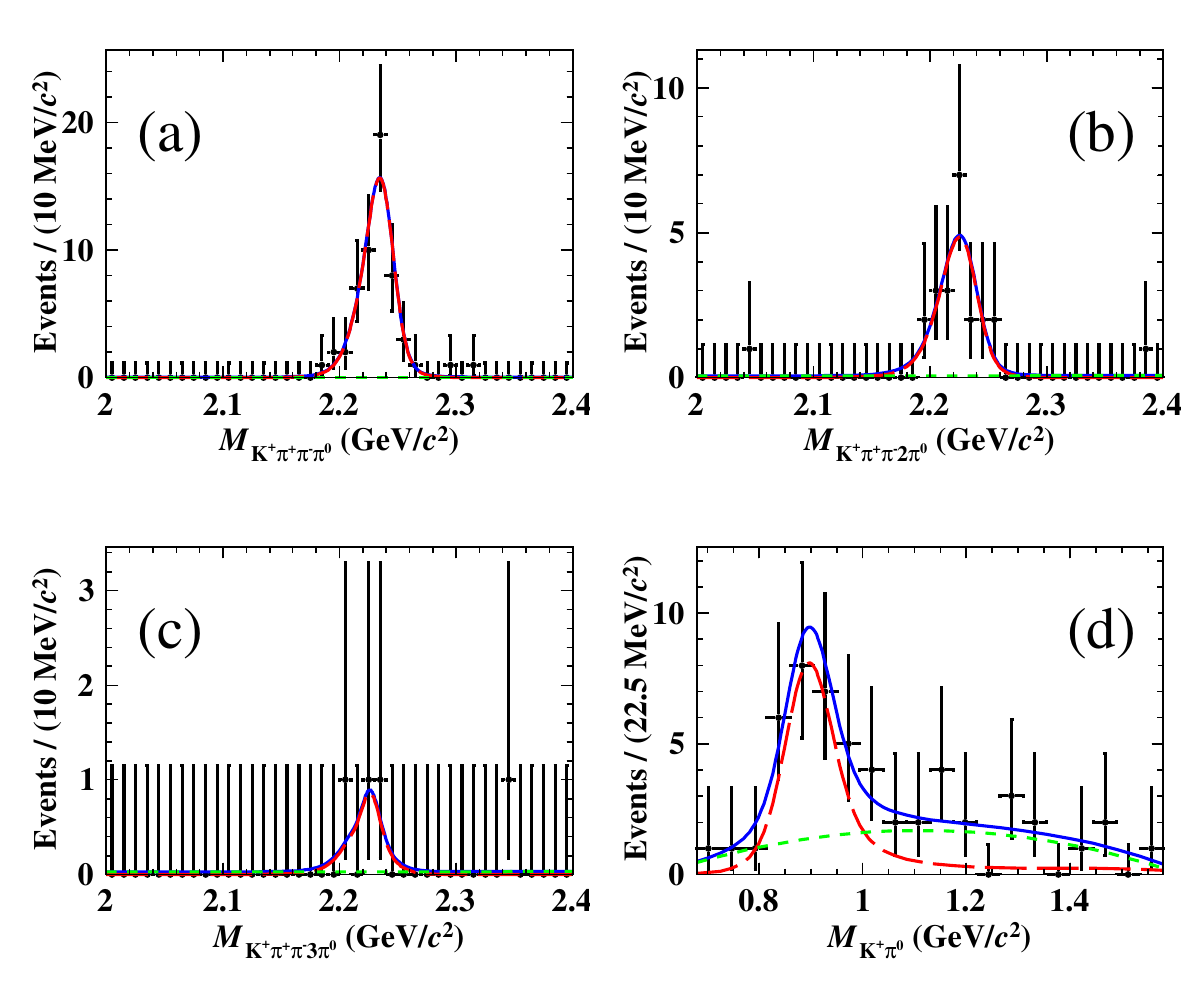}
  \caption{%
     Distributions of the $M_{K^+ \pi^+ \pi^- k\pi^0}$ (a)-(c) and $M_{K^+\pi^0}$ (d) in data (black dots with error bars).
  }
  \label{fig:annihilation}
\end{figure}

\section{Outlook and summary}

The measurements described above demonstrate a new use of
high-luminosity $e^+e^-$ experiments: short-lived hyperons produced in
charmonium decays can serve as incident particles for scattering on
the detector material.
This approach provides access to both hyperon--nucleon and
antihyperon--nucleon interactions within the same experimental
environment.

The program is already being extended beyond the first measurements.
BESIII has studied inelastic $\Lambda$--nucleus interactions,
including $\Lambda N\to\Sigma^+X$~\cite{BESIII2024LambdaN},
and more recent measurements include
$\Sigma^+n\to\Lambda p$, $\Sigma^+n\to\Sigma^0p$, and
additional $\Xi N$ reactions~\cite{BESIII2026Sigma,BESIII2026Xi}.
These channels progressively enlarge the experimental information
available for the strange sector of baryon--baryon forces.

With the large existing charmonium samples at BESIII, further channels
and more differential measurements are expected.
In the longer term, substantially larger charmonium samples at future
facilities such as the Super Tau-Charm Facility would further improve
the precision and extend the range of accessible hyperon--nucleon
reactions.
Together, these measurements establish detector-material scattering
as a complementary experimental approach to the study of the strange
nuclear force.

\section*{Acknowledgments}

The author acknowledges support from the U.S. Department of Energy under Contract No. DE-FG02-05ER41374.


\end{document}